\documentclass[11pt]{article}

\usepackage{./mystyle}

\title{A BPHZ Theorem in Configuration Space}
\author{Steffen Pottel \thanks{Electronic address: steffen.pottel'at'yahoo.de}}
\affil{\small K\"uhne Logistics University, 20457 Hamburg, Germany }
\date{\today}

\begin{document}

\maketitle

\abstract{\begin{center}\begin{minipage}{0.8\textwidth}
	\noindent
	The concept of BPHZ renormalization is translated into configuration space.
  After deriving the counterpart for the regularizing Taylor subtraction, a new version of Zimmermann's convergence theorem by means of the forest formula is proved.
  Furthermore, a sufficient condition on the algebraic decay of the integrand is formulated such that the constant coupling limit exists in the new setting. 
	\end{minipage}\end{center}
}

%\keywords{Weighted Graphs; Renormalization; BPHZ; Configuration Space}

%\ccode{Mathematics Subject Classification 2000: 81T15, 81T18}

% change in layout
\setlength{\parindent}{0em}
\setlength{\parskip}{1.0ex plus 0.5ex minus 0.2ex}

\section{Introduction}

In the conventional approach to quantum field theory, correlation functions are computed using an argument from perturbation theory, i.e. the correlation functions are expressed in terms of weighted Feynman graphs and computed in momentum space after Fourier transformation. 
In general, those weights contain diverging integrals over free variables and require a well-defined constructive prescription in order to render them finite up to certain ambiguities, which get fixed by employing suitable normalization conditions. 
Such a prescription is called renormalization scheme, if it fulfills additionally physically reasonable properties among which are unitarity, covariance and causality. 
The Bogoliubov-Parasiuk-Hepp-Zimmermann (BPHZ) scheme is based on the idea of Bogoliubov and Parasiuk \cite{Bogoliubov:1957gp} to use a variation of the Hadamard regularization of singular integrals in the subtraction of divergent contributions. 
Usually, Hadamard regularization consists of subtracting Taylor polynomials of the function about the singular points to a sufficient order such that the remainder becomes integrable.
Since quantum fields are distributions, the Taylor subtractions have to be carried out on test functions in general, which is the method chosen in Epstein-Glaser renormalization \cite{Epstein:1973gw}.
However, the objects of interest in momentum space are vertex functions, which implies that  test functions are not at our disposal.
Nevertheless there exist external momenta to each integration over a free variable in those vertex functions so that performing the Taylor subtraction with respect to external momenta has the desired regularizing effect.
Bogoliubov and Parasiuk named this the $R$-operation.
It was then rigorously proved by Hepp \cite{Hepp:1966eg} that this prescription leads to a renormalization scheme and since then has been referred to as the BPH scheme. 
We may separate Zimmermann's contribution to BPHZ renormalization into two parts. 
In \cite{Zimmermann:1968mu}, it is shown how to introduce and remove a regularization of the integrals by analytic continuation and, furthermore, sufficient conditions on the existence of the free integrations are given. 
In \cite{Zimmermann:1969jj}, a combinatorial problem stemming from the existence of overlapping divergent integrals, which cannot be treated simultaneously by subtractions, is solved. 
Defining the forest formula for the $R$-operation, any sequence of integrations leads to finite values. 
Indeed, the forest formula can also be applied in other renormalization schemes when one deals with weighted Feynman graphs \cite{Hollands:2010pr,Duetsch:2013xca,Gere:2015qsa} and, depending on the chosen regularization and subtraction, can be proved in a different fashion in comparison to Zimmermann's approach \cite{Collins:1984xc}. 
In \cite{Connes:1999yr}, Connes and Kreimer established that the deeper mathematical structure of weighted graphs and their singularities can be found in the realm of Hopf algebras. 
This fact has recently caught attention in stochastic analysis \cite{hairer2016analyst} using the language of regularity structures. 
However, neither of these works employ the $R$-operation as regularizing element.\\
The result of the present work is the version of Zimmermann's convergence theorem \cite{Zimmermann:1969jj} in configuration space, which is developed using all elements of the original formulation of BPHZ renormalization and constitutes a generalization of the BPHZ prescriptions in \cite[Section 10.3]{Steinmann:2000nr} and \cite{Feldman:1985th}.
By Fourier transformation, the $R$-operation can be transferred to configuration space, where the singularities of large momenta occur in weights when (sub-)graphs are contracted to a point. 
In contrast to the momentum space version, one cannot view the weights as functions over configuration space in general. 
Defining a renormalization scheme then, among other things, turns into the task of extending the weights as distributions to the whole space, i.e. including configurations of contracted (sub-)graphs. 
Indeed, the condition on the extendability of distributions \cite[Chapter 3]{hormander1990analysis} can be reformulated into a condition on local integrability of functions assuming sufficient regularity of the weights. 
It is shown below that graph weights meet this condition after the configuration space version of the $R$-operation is applied to them.
The main difference to approaches mentioned above, which use a different regularization technique, lies in the chosen subtraction prescription and the classification of singular components resulting from it.
The latter refers to the additional subtractions of subdivergencies in the BPHZ scheme, which do not have to be performed in other schemes, for instance, dimensional regularization.
There is an analogous observation in the configuration space formulation. 
While the present approach does not require any additional structures apart from the graph weight, it seemingly bears the drawback that the improvement of the integrand in question also worsens the singular behavior of the complementary weight. 
However, it turns out that this is not harmful since the overall singular behavior of the graph weight remains unchanged. \\
In view of applications in quantum field theory, local integrability is not sufficient in the transition to constant couplings \cite{Epstein:1975gp}. 
The problem is of particular interest for massless quantum fields when the edge weights, associated to fundamental solutions of the wave equation, have only algebraic decay for large distances. 
In \cite{Lowenstein:1975rf,Lowenstein:1975ps}, Lowenstein and Zimmermann modify the BPHZ scheme and show that the modified $R$-operation does not introduce new singularities in the weight at small momenta. 
We transfer their result into the configuration space approach, where, without a modification of the $R$-operation and under certain assumptions on the algebraic decay, the graph weights are absolutely integrable.
This constitutes a significant advantage compared to the momentum space version since massive and massless theories can be treated on equal footing. \\
The paper is structured as follows.
In Section \ref{se:statement} we derive the configuration space version of the $R$-operation and the forest formula.
The result on local integrability is proved in Section \ref{se:L1loc}.
Afterwards absolute integrability of the graph weights is shown in Section \ref{se:L1}.
Finally, the paper is concluded in Section \ref{se:conlusion}.

\section{Statement of the result}\label{se:statement}

The convergence theorem of Zimmermann \cite{Zimmermann:1969jj} is proved for so-called Feynman integrals in momentum space, which we denote by
\begin{align}\label{eq:pFeynmanIntegral}
  \int_{\IR^{md}}\limits dk\, \hat u(p,k) \quad \mbox{ with } \quad p \in \IR^{nd} \, ,
\end{align}
where $d\in \IN$ is the dimension of the underlying space and $m,n\in\IN$ are the number of free integration variables and the number physical momenta, respectively.
Those integrals can be represented as weights over graphs, better known as Feynman graphs, i.e.\ a set of rules assigning an element of the graph to an analytical quantity of a quantum field theory \cite{Itzykson:1980rh}. 
It is this link to discrete structures, which turns out to be helpful for tracking and controlling the combinatorial properties of Feynman integrals in connection with Bogoliubov's $R$-operation.
\begin{definition}
  A graph $\G(V,E)$ is a pair of finite sets, the vertex set $V(\G)$ and the edge set $E(\G)$.
  There exists a map $\pa : E \rightarrow V\times V$, with $\pa e = (s(e),t(e))$ defining an orientation of $\G$ for $s,t:E\rightarrow V$. 
  If there exists two edges $e$, $e'\in E$ with $\{s(e),t(e)\}=\{s(e'),t(e')\}$, we refer to $\G(V,E)$ as multigraph.
  Otherwise $\G(V,E)$ is called simple.\\
  Furthermore, a weighted graph is a graph $\G(V,E)$ together with the assignment of a numerical value to each element in $V(\G)$ and $E(\G)$, respectively.
\end{definition}
We may emphasize the relation of \eqref{eq:pFeynmanIntegral} to graphs by writing explicitly $\hat u(p,k) = \hat u[\G](p,k)$ with $\G = \G(V,E)$ and break this down even further to vertex weights $\hat u[v]$, $v\in V(\G)$, and edge weights $\hat u[e]$, $e\in E(\G)$, namely
\begin{align}
  \hat u[\G] = \prod_{v\in V(\G)} \hat u[v] \prod_{e\in E(\G)} \hat u[e] \, .
\end{align}
Next, we resolve the origin of the integrations.
Each edge is associated to Green functions of a linear differential operator, which can usually be written as rational functions in momentum space.
Hence an individual momentum variable $p_e$ is assigned to each edge $e\in E(\G)$.
However, at each vertex with more than one incident edge, momentum conservation is enforced.
This leads to an ambiguity in the assignment of momenta if the graph contains cycles, i.e.\ if there exists two disjoint paths in the graph connecting a pair of vertices, and results in an integration over a free momentum variable in that cycle.
With this information, it is straightforward to derive the corresponding quantity in configuration space.
After Fourier transformation, the graph weights become constants $u[v] = \mbox{const.}$ for $v\in V(\G)$ and Green functions $u[e] = G(x_{s(e)}, x_{t(e)})$ for $e\in E(\G)$, respectively.
We recall that Green functions are singular for coinciding arguments, i.e. $x_{s(e)} = x_{t(e)}$, so that for edges connecting the same pair of vertices in a multigraph, the singular points of their weights coincide.
Hence without loss of generality we may consider only simple graphs with edge weights $u[e](x_{s(e)},x_{t(e)}) \in C^\infty(\IR^{2d} \setminus \{ (x_{s(e)},x_{t(e)})\in \IR^{2d} \mid x_{s(e)} = x_{t(e)} \} )$ throughout the document.
Further we may assign weights $u[v] \in C^\infty(\IR^d)$ to every $v \in V(\G)$. 
\begin{definition}
  Let $\G(V,E)$ be a simple graph and denote by $x_v$ the location of the vertex $v\in V(\G)$ in configuration space. 
  We define the large graph diagonal by
  \begin{align}
    \oo & \bydef \{x\in \IR^{d\lvert V(\G)\rvert} \mid \exists \g\subset\G \mbox{ connected } \forall v,w\in V(\g), v\neq w: x_v=x_w \} \\
    \intertext{and the thin graph diagonal by}
    \OO & \bydef \{ x \in \IR^{d\lvert V(\G)\rvert} \mid \forall v,w\in V(\G): x_v=x_w \}.                
  \end{align}
\end{definition}
With this definition, we have
\begin{align}
  & u[\G] = \prod_{e\in E(\G)} u[e] \prod_{v\in V(\G)} u[v] \in C^\infty(\IR^{d\lvert V(\G)\rvert}\setminus\oo), \\
  \intertext{where for each $e\in E(\G)$ and $v\in V(\G)$}
  & u[e]\in C^\infty(\IR^{2d}\setminus \OO) \quad \mbox{and} \quad u[v]\in C^\infty(\IR^d)  
\end{align}
holds, respectively. 
It is worth emphasizing that the vertex weights $u[v]$ with $v\in V(\G)$ should not be considered as test functions in the dual space of some distribution but rather in analogy to Wick powers of some scalar quantum field \cite{Hollands:2001fb}.
In the same sense, we may think of $u[\G]$ as a distributional kernel, which is currently only defined on $\IR^{d\lvert V(\G)\rvert}\setminus\oo$ and shall be extended to $\IR^{d\lvert V(\G)\rvert}$.
By the Hahn-Banach theorem, this extension exists and is unique if $u[\G]$ is locally integrable.
In order to examine the integrability of $u[\G]$, we need to control the scaling of graph weights near the large graph diagonal.
The following notion was first introduced by Steinmann for distributions in \cite{steinmann1971perturbation}.
\begin{definition}\label{de:UVScaling}
  Let $u\in C^\infty(\IR^{nd}\setminus\OO)$. Then the UV-scaling degree of $u$ is defined as
  \begin{align}
    \uvs(u) \bydef \inf\left\{\al\in\IR \mid \lim_{\la\rightarrow0}\limits \la^\al u(\la x) = 0 ; x\in\IR^{nd}\setminus\OO \right\}. 
  \end{align}
\end{definition}
Translating this definition to graph weights, the UV-scaling degree of $u[\g]\in C^\infty(\IR^{d\lvert V(\g)\rvert}\setminus\OO)$ measures the behavior $u[\g]$, if $\g$ is contracted to a point.
For our result, it turns out to be more reasonable to work with the \emph{UV-degree of divergence}
\begin{align}
    \uvd(u[\g]) \bydef \uvs(u[\g]) - d(\lvert V(\g)\rvert-1)
\end{align}
of a (sub-)graph $\g\subseteq\G$, which relates the scaling degree of $u[\g]$ with the dimension of the integration, which is required to meet the thin graph diagonal of $\g$. 
With this, $u\in C^\infty(\IR^{nd}\setminus\OO)$ is locally integrable if $\uvd(u) < 0$.
Note that the scaling acts on vertices but not on lines such that the scaling degree takes all edges connecting the scaled vertices into account. 
We refer to those type of graphs as \emph{full vertex parts}. 
Furthermore, we observe that there exist elements in the large graph diagonal, which overlap in the sense that for two graphs $\g\subset\G$ and $\g'\subset\G$ graph contractions cannot be performed independently of each other. 
\begin{definition}
  Two graphs $\g$ and $\g'$ are overlapping, denoted by $\g \olap \g'$, if none of the following conditions hold
  \begin{align}\label{eq:non_overlapping_conditions}
    V(\g)\subseteq V(\g'), \quad V(\g)\supseteq V(\g'), \quad V(\g)\cap V(\g') = \emptyset \,.
  \end{align}
  Otherwise they are non-overlapping, denoted by $\g\nolap\g'$.
\end{definition}
Subgraph weights with positive degree of divergence and overlapping vertex sets pose the major problem apart from employing a proper method of reducing the degree of divergence constructively. 
For the former, Zimmermann introduced in \cite{Zimmermann:1969jj} the notion of forests, which consists of all sets of non-overlapping graphs $\g\subseteq\G$.
\begin{definition}
  A $\G$-forest $F$ is a partially ordered set (poset) over $V(\G)$, where elements in $F$ are ordered by usual set inclusion $\subseteq$ and one relation of \eqref{eq:non_overlapping_conditions} holds for each pair in $F$.
\end{definition}
For the reduction of the degree of divergence, the BPHZ method uses Taylor polynomials for the manipulation of the graph weights. 
We denote such polynomials for a sufficiently smooth function $f$ by
\begin{align}
    t^{k}_{x\mid\ol{x}} f(x) \bydef \sum_{\lvert\al\rvert=0}^{k}\limits \frac{(x-\ol{x})^\al}{\al!} f^{(\al)}(\ol x),
\end{align}
where $\ol x$ is the point about which the subtraction is performed and $\al\in\IN^d$ is a multiindex indicating the derivatives of $f$. 
We choose the point of subtraction $x_{\ol \g}$ to be located at the thin graph diagonal of the to-be-renormalized (sub-)graph $\g$. 
For any graph $\g\subseteq\G$, its thin graph diagonal depends on the configuration of $\g$ in space, i.e. on $x_v\in\IR^d$ for $v\in V(\g)$, and by this the point of subtraction is not a constant but changes according to the configuration of the graph. 
We set $\ol v\in V(\ol{\g})$ to be the vertex which location is computed by
\begin{align}\label{eq:CMCoordinate}
  x_{\ol \g} \bydef \sum_{v\in V(\g)} \xi_v x_{v}, 
\end{align}
where 
\begin{align}
  0 < \xi_v \leq 1 \quad \mbox{ and } \quad \sum_{v\in V(\g)} \xi_v = 1.
\end{align}
We note that the choice of $x_{\ol \g}$ is admissible since the subtraction point coincides with the thin diagonal, i.e. $x_{\ol \g} = x_v$ for $v \in V(\g)$ if $x_v=v_w$ holds for all $w \in {V(\g)\setminus\{v\}}$.
Thus it generalizes the center of mass subtraction point $\ol{x} = \lvert V(\g)\rvert^{-1} \sum_{v\in V(\g)} x_v$ in \cite{Steinmann:2000nr} and the fixed vertex subtraction point $\ol{x} = x_v$ for a fixed $v\in V(\g)$ in \cite{Feldman:1985th}, where in the latter case subtractions are performed on test functions instead of the graph weight.\\
Furthermore, we observe that the action of a Taylor operator $t^d_{x_\g\mid x_{\ol \g} }$ on $u[\g]$, where $d>0$ and $x_\g = \{x_v\lvert v\in V(\g)\}$, is ill-defined by definition of the edge weights $u[e]$ with $e\in E(\g)$.
However, the incident lines of $V(\ol{\g})$, i.e. after $\g$ is contracted to a point, are smooth in a neighborhood of $x_{\ol \g}$ so that Taylor polynomials may be computed on those.
\begin{definition}
  Let $\G$ and $\g\subset\G$ be graphs with weights $u[\G]$ and $u[\g]$, respectively. 
  We denote by
  \begin{align}
    t(\g) \bydef t^{d(\g)}_{x_\g \mid x_{\ol\g}} \IP(\g)
  \end{align}
  the Taylor operator for the graph weight $u[\g]$, where
  \begin{align}
    d(\g) & \bydef \lfloor \uvd(u[\g]) \rfloor \,,\\
    t(\g) & = 1 \quad \mbox{for} \quad d(\g)<0 \,, \\
    x_\g & = \{x_v\lvert V\in V(\g)\}
  \end{align}
  and the action of the operator $\IP(\g)$ is given by
  \begin{align}
    t^{d(\g)}_{x_\g\mid x_{\ol\g}} \IP(\g) u[\G] = u[\g] \, t^{d(\g)}_{x_\g\mid x_{\ol\g}} \, u[\G\lineco\g] \, .
  \end{align} 
  Here, the set difference $\lineco$ is meant to be computed with respect to the set of lines $E$, i.e. $\G\lineco\g$ is not a full vertex part and $V(\G\lineco\g)\cap V(\g) \neq \emptyset$. 
\end{definition}
With this the configuration space formulation of the forest formula is as follows.
\begin{definition}
  The $R$-operation on the graph weight $u[\G]$ is given by
    \begin{align}\label{eq:forestformula}
      Ru[\G] \bydef \sum_{F\in\SF}\limits \prod_{\g\in F} (-t(\g)) u[\G], 
    \end{align}
    where $\SF$ is the set of all $\G$-forests and the Taylor operators are ordered in the sense that $t(\g)$ appears left of $t(\g')$ if $\g\supset \g'$ and no order is preferred if $V(\g) \cap V(\g') = \emptyset$.
\end{definition}
With all necessary notions at hand, we return to the problem of local integrability, which has a positive solution due to the forest formula.
\begin{theorem}\label{th:L1loc}
  Let $u[\G]\in\ C^\infty(\IR^{d\lvert V(\G)\rvert}\setminus\oo)$ be the weight over a graph $\G$, which has positive scaling degree at the large graph diagonal. 
  Then 
  \begin{align}
    Ru[\G] \in L^1_\mathrm{loc}(\IR^{d\lvert V(\G)\rvert})\,.
  \end{align}
\end{theorem}
In relation to the momentum space scheme, one may additionally ask for global integrability with respect to internal vertices, i.e. vertices with more than one incident edge. 
This corresponds to the transition of coupling functions with compact support to coupling constants, which is indeed the setting of Zimmermann's result. 
We may be able to control such a limit of constant couplings by examining the influence of the $R$-operation on the long range behavior of the weights.
\begin{definition}
  Let $u\in C^\infty(\IR^n)$. Then the IR-scaling degree of $u$ is defined by
  \begin{align}
    \irs(u) \bydef \sup \left\{ \al\in\IR \mid \lim_{\La\rightarrow \infty} \La^\al u(\La x) = 0 \right\}
  \end{align} 
  and the IR-degree of divergence is given by
  \begin{align}
    \ird(u) \bydef \irs(u) - n.
  \end{align}
\end{definition}
For short ranges, the external lines of the considered subgraph are smooth at the point of coincidence and only internal lines determine the local integrability of the weight. 
Instead, for large arguments, the external lines contribute with their decay behavior. 
This was already exploited for the momentum space scheme in \cite{Lowenstein:1975ps}.
Therefore we consider the edges of the full vertex parts in the UV, but all incident edges of the involved vertex set in the IR. 
In order to distinguish both notions, a graph involving all incident edges of a given vertex set $V(\g)$ is denoted by $\dot\g$. 
However, this is implicit for the IR-scaling degree such that we still write $\irs(u[\g])$.
\begin{theorem}\label{th:L1}
  Let $u[\G]\in C^\infty(\IR^{d\lvert V(\G)\rvert}\setminus\oo)$ be the weight over $\G$, which has positive scaling degree at the large graph diagonal. Suppose that the IR-degree of divergence is positive for all $u[\g]$, $\g\subseteq\G$. Then
  \begin{align}
    Ru[\G] \in L^1(\IR^{d \lvert I\rvert})
  \end{align} 
  for any $I\subset V(\G)$.
\end{theorem}
We observe that the $R$-operation remains the same for graph weights with algebraic decay, which, for instance, can be found in quantum field theories with massless particles.
In the momentum space version of the BPHZ scheme, the treatment of massless particles is only possible after the introduction of an auxiliary mass term, which has to be included in the Taylor subtractions \cite{Lowenstein:1975ps}.
However this bond disappears in the configuration space formulation.

\section{Proof of Theorem \ref{th:L1loc}}\label{se:L1loc}

We want to show that, for each forest $F$, the UV-degree of divergence of $\prod_{\g\in F} (-t(\g)) u[\G]$ is below a certain threshold in reference to a chosen integration over vertices $I\subseteq V(\G)$. As a rough estimation, we know that the empty forest is an element in the set of all forests $\SF$. Knowing that there exists a $\g\subset \G$ which has $d(\g)\geq0$, we cannot expect for all integrations to find better behavior than the one of the empty set. Analogously to Zimmermann \cite{Zimmermann:1969jj}, a reordering of the forest formula is feasible, but requires a careful treatment due to the relation between graph contractions and overlap. The proof is performed in two steps. After reordering the elements in forests, we show that the recursive structure of the $R$-operation leads to the desired scaling of the weight at the large graph diagonal.

\subsection{Reordering}

Our strategy for reordering the forest formula is inspired by the proof of Zimmermann in \cite{Zimmermann:1969jj}, i.e. for any chosen integration set $I$, the forests are reordered such that we meet the condition on the degree of divergence. As a first step, we characterize graphs.
\begin{definition}\label{de:graphchar}
  Consider a subgraph $\g\subseteq\G$ and an integration set $I_m\subseteq V(\G)$ with $\lvert I_m\rvert =m$. Then $\g$ is called 
  \begin{enumerate}
    \item \emph{variable} with respect to (wrt) $I_m$ if $\lvert V(\g) \cap I_m\rvert =\lvert V(\g)\rvert-1$,
    \item \emph{integrated} wrt $I_m$ if $\lvert V(\g) \cap I_m\rvert=\lvert V(\g)\rvert$,
    \item \emph{constant} wrt $I_m$ otherwise.
  \end{enumerate}
\end{definition}    
\begin{remark}\label{rm:intconst}
  Let $\g\subset\G$ and $\g'\subset\g\subset\g''$. Then
  \begin{enumerate}
    \item $\g'$ is integrated if $\g$ is integrated,
    \item $\g''$ is constant if $\g$ is constant,
    \item $\g'$ is variable or integrated if $\g$ is variable.
  \end{enumerate}
\end{remark}
Let $A$ be a finite set, $I\subseteq A$ and $F$ be a forest over $A$, i.e. a poset over $A$ such that each pair of elements in $F$ fulfills a relation of \eqref{eq:non_overlapping_conditions}.
Further consider any $f\in F$ with maximal subelements $f_1,...f_a$ with $f_j\in F$, where $f_j\subset f$ is maximal, if there exists no $f_i\subset f$ with $i\neq j$ and $f_i\in F$ such that $f_i\supset f_j$. We set the following rules for reduction of subelements. If $f_j$ is variable, we reduce $f_j$ to a constant vertex. If $f_j$ is integrated, we reduce $f_j$ to an integrated vertex. If $f_j$ is constant, we take the set difference of $f_j$ with $f$ itself. Without loss of generality the elements $f_1,...,f_b$ with $b\leq a$ are either integrated or variable and the reduced elements are given by
\begin{align}
  \tilde f \bydef (f/f_1...f_b)\setminus f_{b+1}...f_a 
\end{align}
where $f/g$ means the reduction of $g$ inside $f$ to a vertex and $f\setminus g$ is the usual set difference wrt the vertex sets. Then we define the set
\begin{align}
  \tilde F \bydef \{ \tilde f \mid f \in F, \, f_1...f_a \mbox{ reduced}, \, f_j \in F \mbox{ maximal wrt } f \}. 
\end{align}
Recall that $F$ is a forest, where elements $f$ potentially fulfill additional conditions.
In the case of Zimmermann, the set $A$ is the set of lines $E(\G)$ of a graph $\G$ and each $f$ was required to be the lines of a one-particle-irreducible (1PI) (sub-)graph. We are considering the vertex set $V(\G)$ and require elements of forests to be full vertex parts. But our construction holds for more general conditions.
Without specifying it further, let us refer to the condition as $C$.\\
In order to enable us to relate various forests, we take the intersection of $\tilde f$ with the integration set $I$ to determine the integrated elements in $\tilde f$.
\begin{align}\label{eq:redint}
  I_{\tilde f} \bydef \tilde f\cap I 
\end{align}
Note that $I_{\tilde f}$ does not necessarily meet the condition $C$. Nevertheless we construct elements $\tilde g \subseteq \tilde f$, where $\tilde g$ is variable wrt $I$, meets $C$ and is maximal wrt $C$ in the sense that there exists no variable $\tilde g' \subseteq \tilde f$ meeting $C$ such that $\tilde g \subset \tilde g'$. This construction is not unique in general, i.e. following from the definition of variable elements, there may exist overlapping elements. 
Suppose that we can find a configuration of non-overlapping elements so that we define
\begin{align}
  \tilde G_F \bydef \{ \tilde g \subset \tilde f \mid \tilde g \mbox{ variable wrt $I$ \& maximal wrt }C,\, f\in F,\, \tilde g \nolap \tilde g' \, \forall\, \tilde g'\in \tilde G\}. 
\end{align}
From this set, we recover the ``full'' elements by ``blowing up'' the reduced graphs again, i.e.
\begin{multline}
  G_F \bydef \{ g\subset f \mid g \bydef \tilde g \cup (f_{i(1)}...f_{i(c_g)}),
   \tilde g \in \tilde G_F, f\in F, f_{i(j)} \mbox{ max wrt }f, f_{i(j)}\cap \tilde g \neq \emptyset \} .
\end{multline}
Furthermore we need to define two sets that will gain importance at a later stage of the proof. We set 
\begin{align}
  \ol f = f \setminus f_1...f_a ,
\end{align}
define 
\begin{align}
  F' \bydef \{ f\in F \mid \ol f \mbox{ is constant} \} 
\end{align}
together with the set of maximal variable elements inside constant elements
\begin{align}
  H_F \bydef \{ f \in F \mid f \mbox{ variable } \& f \mbox{ maximal element of } f'\in F' \}. 
\end{align}
We begin reordering of the forest formula considering the union of $F$ and $G_F$.
\begin{lemma}
  $(F\cup G_F, \subseteq)$ is a forest.
\end{lemma}
\begin{proof}
  Consider $g_f,g'_f, g_{f'}\in G_F$. By definition $g_f= g'_f$ or $g_f \cap g'_f =\emptyset$ holds. For $f\cap f'=\emptyset$, $g_f\cap g_{f'}=\emptyset$ holds, and for $f\subset f'$, we refer again to the definition of $G_F$ from which $g_f\cap g_{f'}=\emptyset$ follows by Remark \ref{rm:intconst}.  \\
  Next, take $f,f'\in F$ and $g_f\in G_F$. Then
  \begin{align}
    f \cap f' =\emptyset & \Rightarrow g_f\cap f'=\emptyset, \\
    f \subset f'         & \Rightarrow g_f \subset f', \\
    f \supset f'         & \Rightarrow f'\subset f_j \mbox{ maximal wrt $f$} \\
                         & \Rightarrow \mbox{either } f'\cap g_f=\emptyset \mbox{ or } f'\subset g_f .
  \end{align}
  Since $(F,\subseteq)$ is a forest, so is $(F\cup G_F,\subseteq)$ as asserted.
\end{proof}
Zimmermann introduced the notion of "complete" forests in his proof. This is not a good choice for our approach since all forests are complete posets due to the finiteness of $A$.
Instead, we introduce another notion, which expresses the same idea.
\begin{definition}
  A forest $F$ over a finite set $A$ is \emph{saturated with respect to $I$}, if for every $f\in F$ either $\ol f \cap I = \emptyset$ or $\ol f$ is variable or integrated.
\end{definition}
\begin{lemma}\label{le:saturation}
  $(F\cup G_F,\subseteq)$ is a saturated forest.
\end{lemma}
\begin{proof}
  Let $f\in F'$. Then $\ol f \cap I = \emptyset$, where $\ol f$ is computed wrt $F\cup G_F$, since $f$ only has variable or constant subgraphs by construction. For $f\in F\setminus F'$, $\ol f$ is either variable or integrated. If $g\in G_F$, then $\tilde g$ is variable by construction and hence $\ol g$ is either variable or integrated. This proves the assertion.
\end{proof}
From now on we use the notation $S_I(F)$ for the saturation of $F$ with respect to $I$. 
\begin{definition}\label{de:base}
  Let $(F,\subseteq)$ be a forest. The base $B_I(F)$ of $F$ is given by the relation
  \begin{align}
      & B_I(F) \bydef F\setminus H_F \setminus A \\
      & B'_I(F) \bydef B_I(F) \cup A.              
  \end{align}
\end{definition} 
From Definition \ref{de:base} and Lemma \ref{le:saturation}, it is clear that we can always find base $B_I(F)$ and saturation $S_I(F)$ given a forest $(F,\subseteq)$.
We want to show that $B_I(F)$ and $S_I(F)$ are directly related.
\begin{lemma}
  Let $(F,\subseteq)$ be a forest with base $B_I(F)$ and saturation $S_I(F)$. Then $B'_I(S_I(F)) = B'_I(F)$ and $S_I(B'_I(F))=S_I(F)$.
\end{lemma}
\begin{proof}
  By definition, we have 
  \begin{align} 
    B'_I(S_I(F)) = (F\cup G_F)\setminus H_{F\cup G_F}. 
  \end{align}
  Then $B'_I(S_I(F))=B'_I(F)$, if $H_{F\cup G_F}=G_F\cup H_F$, since 
  \begin{align}
    (F\cup G_F)\setminus H_{F\cup G_F} = (F\cup G_F)\setminus (G_F\cup H_F) = F \setminus H_F = B'_I(F). 
  \end{align}
  But the latter only holds, if $(F\setminus H_F)'=F'$, which follows from
  \begin{align}
    f\in H_F                        & \Rightarrow f\notin F',\\
    f\in F\setminus H_F,\, f\in F'  & \Rightarrow \ol f^F \cap I \subseteq \ol f^{F\setminus H_F} \cap I \\
                                    & \Rightarrow f \in (F\setminus H_F)',\\
    f\in F\setminus H_F, f\notin F' & \Rightarrow \ol f^F \mbox{ is variable or integrated} \\
                                    & \Rightarrow f_1,...,f_a \mbox{ maximal wrt } f \mbox{are not in } H_F \\
                                    & \Rightarrow f\notin (F\setminus H_F)'.
  \end{align}
  For 
  \begin{align}
    S_I(B'_I(F)) \bydef (F\setminus H_F) \cup G_{F\setminus H_F}, 
  \end{align}
  we obtain the assertion, if $G_{F\setminus H_F}= G_F \cup H_F$, since 
  \begin{align}
    (F\setminus H_F) \cup G_{F\setminus H_F} = (F\setminus H_F) \cup G_F \cup H_F = F \cup G_F = S_I(F). 
  \end{align}
  We already know that $(F\setminus H_F)'=F'$. Hence $G_{F\setminus H_F}= G_F \cup H_F$ follows from 
  \begin{align}
    \ol f^F\setminus \ol g^F_f = \ol f^{F\setminus H_F} \setminus \ol g^{F\setminus H_F}_f 
  \end{align}
  for $f\in F'=(F\setminus H_F)'$. Certainly, 
  \begin{align}
    \ol f^F\setminus \ol g^F_f \subseteq \ol f^{F\setminus H_F} \setminus \ol g^{F\setminus H_F}_f 
  \end{align}
  holds. If there exists an $f_0\in H_F$ with $f_0\subset f$, then $f_0\in G_{F\setminus H_F}$, and otherwise we have  
  \begin{align}
    \ol f^{F\setminus H_F} = \ol f^F. 
  \end{align}
  In both cases it follows 
  \begin{align}
    (\ol f^{F\setminus H_F} \setminus \ol g^{F\setminus H_F}_f) \setminus  (\ol f^F\setminus \ol g^F_f) = \emptyset. 
  \end{align}
  This proves the assertion. 
\end{proof}
\begin{proposition}\label{pr:equiclass}
  Let $(S,\subseteq)$ be a saturated forest over $A$ with base $B$. The set $\SF$ of forests $(F,\subseteq)$ with saturation $S$ is given by the condition
  \begin{align}
    B\subseteq F \subseteq S .
  \end{align} 
\end{proposition}
\begin{proof}
  Let $B\subseteq F \subseteq S$. We know that $S_I(B)=S$ and relate $G_B$ and $G_F$ by 
  \begin{align}
      & f\in B' \Rightarrow f\in S' \Rightarrow f\in F, \\
      & f\in B, f\in F' \Rightarrow f\in S' \Rightarrow f\in F' \mbox{ since } \ol f^F\subseteq \ol f^B, \\
      & f\in F\setminus B \subseteq G_B \Rightarrow \ol f^F = \ol f^B \mbox{ by definition of } G_F .
  \end{align}
  With this we obtain $B'=F'$ and $G_F=G_B\setminus(F\setminus B)$. Then 
  \begin{align}
    S_I(F) = F\cup G_F = F \cup G_B\setminus (F\setminus B) = B \cup G_B = S. 
  \end{align}
\end{proof}
Since the result on equivalence classes of forests is established, we are able to perform the reordering of the forest formula. 
\begin{lemma}
  Provided that the equivalence class of a given forest $F$ over $\G$ is given by $B\subseteq F \subseteq S$, the (truncated) weight $Ru[\G]$ over the graph $\G$ is given by
  \begin{align}
    Ru[\G] = \sum_{S\in \SCS} \prod_{\g\in S} (\ch(\g)\IP(\g)) u[\G] 
  \end{align}
  with $\ch(\g)=(1-t(\g))$ for $\g\in H_S$ and $\ch(\g)=-t(\g)$ if $\g\notin H_S$. $\SCS$ denotes the set of all saturated forests.
\end{lemma}
\begin{proof}
  In a first step we write
  \begin{align}
    \sum_{F\in\SF}\limits \prod_{\g\in F} (-t^{d(\g)}_{x_\g\mid x_{\ol{\g}} } \IP(\g)) = \sum_{S\in \SCS} \sum_{B\subseteq F \subseteq S} \prod_{\g\in F} (-t^{d(\g)}_{x_\g\mid x_{\ol{\g}} } \IP(\g)) 
  \end{align}
  We observe that $S\setminus B = (B \cup H_S) \setminus B = H_S$, such that $\g\in F$ is either in $H_S$ or $B$. Now if $\g\in H_S$ then there exists an $F_0$ such that $g\notin F_0$ and likewise if $\g\in B$ then there exist no $F_0$ such that $\g\notin F_0$ since $B\subseteq F_0$ holds. Hence for $\g\in H_S$ we may split the sum into a set of forests containing $\g$ and into set of forests not containing $\g$. Like this we obtain the factor $(1-t(\g))$. Since there exist no forest $F_0$ that does not contain $\g\in B$, the only factor one obtains is $-t(\g)$.  
\end{proof}

\subsection{Recursion}

In the following, we describe the actual recursive structure of the $R$-operation. Suppose $F$ is saturated wrt $I$ and $\ch(.)$ is given as above. For some $\g\in F$, the weight $u[\g]$ after the $R$-operation reads
\begin{align}
  R'_F u[\g] = u[\g_1]...u[\g_a] \prod_{j=1}^a \ch(\g_j) u[\G\lineco \g_1...\g_a \mid_\g] ,
\end{align}
where we denote by $R'_F$ the restricted action of $R$ to elements $\g_j\in F$ which are maximal wrt $\g$. Note that by construction, all $\g_j$ are mutually disjoint. Further we recall that we have to scale every vertex that is in $I$, i.e. $x_v\mapsto \la x_v$ if $v\in I$. Since we just picked any $\g$, it suffices to look at $I_\g$ from \eqref{eq:redint}. With the help of Definition \ref{de:UVScaling}, we are able to examine the effect of Taylor operators on the graph weights. For simplicity we consider first sufficiently smooth functions of several real variables. Then we have for a family of functions $f_1,...,f_n$ \cite{Lowenstein:1975rg}
\begin{align}
  \uvs \Big( \sum_{i=1}^n f_i \Big)  & \leq \max_i \{\uvs(f_i)\} \label{eq:LZUVProd}\\
  \uvs \Big( \prod_{i=1}^n f_i \Big) & = \sum_{i=1}^n \uvs(f_i)  \label{eq:LZUVSum}.
\end{align}
Since we want to analyze Taylor polynomials of these functions, we make use of the following \cite{Brunetti:1999jn}.
\begin{lemma}\label{le:BFscale}
  For $\al$ multiindex, a smooth function $f\in \IR^n\setminus\OO$ we have
  \begin{align}
    \uvs(x^\al f)   & \leq \uvs(f) - \lvert\al\rvert, \label{eq:scalingmoments} \\
    \uvs(\pa^\al u) & \leq \uvs(f) + \lvert\al\rvert. \label{eq:scalingderivatives}      
  \end{align}
\end{lemma}
\begin{lemma}\label{le:uvstaylor}
  Let $f:\IR^n\times \IR^m \rightarrow \IC$ and $k$-times continuously differentiable in the first entry. Let further $t^d_{x\mid\ol x}$ be the Taylor operator and $d\leq k$. Then
  \begin{align}
      & \uvs_x(t^d_{x\mid\ol x} f(x,y)) \leq \uvs_x (f(x,y))       ,                \\
      & \uvs_{x,y}(t^d_{x\mid\ol x} f(x,y)) \leq \uvs_{x,y} (f(x,y))      ,         \\
      & \uvs_{x}((1-t^d_{x\mid\ol x}) f(x,y)) \leq \uvs_x (f^{(d+1)}(x,y)) - d - 1 ,
  \end{align}
  where $\uvs_z$ denotes the scaling degree where only the variable $z$ is scaled.
\end{lemma}
\begin{proof}
  Note that 
  \begin{align}
    \uvs_x(t^d_{x\mid\ol x} f(x,y)) & = \uvs_x\Big(\sum_{\lvert\al\rvert=0}^d \frac{(x-\bar x)^\al}{\al!} f^{(\al)}(\bar x,y) \Big) \\
                                   & = \max_\al \,\uvs_x\left( \frac{(x-\bar x)^\al}{\al!} f^{(\al)}(\bar x,y) \right)        
  \end{align}
  and hence we may fix any $\al$ and apply Lemma \ref{le:BFscale}. 
  First we recall that $f\in C^k(\IR^n)$ and therefore smooth in the scaling limit $\uvs_x$ so that
  \begin{align}
    \uvs_x(f^{(\al)}(\bar x, y)) \leq 0 \stackrel{\eqref{eq:scalingmoments}}{\Rightarrow} \uvs_x\left( (x-\bar x)^\al f^{(\al)}(\bar x,y) \right) \leq -\lvert\al\rvert. 
  \end{align}
  Then we take the maximum over $\al$ and the first assertion follows. 
  Further we observe that $f^{(\al)}(\bar x,y)$ is not a smooth function in the scaling limit with respect to $x$ and $y$. 
  Hence we use
  \begin{align}
    \uvs_{x,y}(f^{(\al)}(\bar x,y)) \stackrel{\eqref{eq:scalingderivatives}}{\leq} \uvs_{x,y}(f(\bar x,y)) + \lvert\al\rvert 
  \end{align}
  and we obtain
  \begin{align}
    \uvs_{x,y}\left( (x-\bar x)^\al f^{(\al)}(\bar x,y) \right) = \uvs_{x,y}(f(\bar x,y)) + \lvert\al\rvert - \lvert\al\rvert = \uvs_{x,y}(f(\bar x,y)), 
  \end{align}
  which is independent of $\al$, so that the second assertion follows. 
  In case of the third assertion, we write
  \begin{align}
    (1-t^d_{x\mid\ol x}) f(x,y) = \sum_{\lvert\al\rvert=d+1} \frac{(d+1)(x- \bar x)^\al}{\al!} %\times \\
    \int (1-\te)^{d+1} f^{(\al)}(\bar x + \te (x-\bar x),y) d\te ,
  \end{align}
  where $f^{(\al)}(\bar x + \te (x-\bar x),y)$ is again a smooth function in the scaling limit with respect to $x$ and $(x - \bar x) \rightarrow 0$ such that the integration becomes independent. Thus we apply Lemma \ref{le:BFscale} for moments and find
  \begin{align}
    \uvs_{x}((1-t^d_{x\mid\ol x}) f(x,y)) \leq \uvs_x (f^{(d+1)}(\bar x,y)) - \lvert\al\rvert = \uvs_x (f^{(d+1)}(\bar x,y))-d-1 
  \end{align}
  This concludes the proof of the lemma.
\end{proof}
With this, we compute the scaling behavior of a graph weight modified by a single Taylor operator. 
\begin{lemma}\label{le:Taylor}
  Let $\G$ be a Feynman graph with vertex set $V$ and edge set $E$ and $u[\G]$ be the weight over $\G$. Then we have
  \begin{align}
      & \uvs_{x_\la} \left( - t^{d(\g)}_{x_\g\mid x_{\ol{\g}}} \IP(\g) u[\G] \right) \leq \uvs_{x_\la} \left( u[\la] \right)                          \\
      & \uvs_{x_\g} \left( \left( 1- t^{d(\g)}_{x_\g\mid x_{\ol{\g}}} \IP(\g) \right) u[\G] \right) \leq \uvs_{x_\g} \left( u[\G] \right) - d(\g) - 1 
  \end{align} 
  for $\la, \g \subseteq \G$, $\g\nolap\la$.
\end{lemma}
\begin{proof}
  We want to apply Lemma \ref{le:uvstaylor} and therefore have to analyze the decomposition of $u[\G]$. Applying the ordering operator $\IP$, we have
  \begin{align}
    - t^{d(\g)}_{x_\g\mid x_{\ol{\g}}} \IP(\g) u[\G] = u[\g] \left(  - t^{d(\g)}_{x_\g\mid x_{\ol{\g}}} u[\G\lineco \g] \right) 
  \end{align}
  and may focus on the second factor. 
  Various cases have to be distinguished according to the relation of the set $V(\g)$ for which Taylor polynomials are computed, and the set $V(\la)$ for which scaling is performed.
  \begin{description}
    \item $V(\la)\subseteq V(\g)$. There is nothing to modify and we write
    \begin{align}
      \uvs_{x_\la} & \left( u[\g] \left(  - t^{d(\g)}_{x_\g\mid x_{\ol{\g}}} u[\G\lineco \g] \right) \right)\\
        & = \uvs_{x_\la}\left( u[\g] \right) + \uvs_{x_\la}\left( - t^{d(\g)}_{x_\g\mid x_{\ol{\g}}} u[\G\lineco \g] \right)        \\
      & \leq \uvs_{x_\la}\left( u[\g] \right) + \uvs_{x_\la}\left( u[\G\lineco \g] \right) \\
      & \leq \uvs_{x_\la}\left( u[\la] \right). 
  \end{align}
    \item $V(\la)\cap V(\g)= \emptyset$. By construction, the point $x_{\ol{\g}}$ has to be disjoint for the coincidence point $x_{\ol{\la}}$. Hence the edges that can be involved in both the Taylor operation on vertices in $V(\g)$ and the scaling on vertices in $V(\la)$ are smooth at the point of coincidence by assumption.
    \item $V(\la)\supset V(\g)$. We can reduce the action of the Taylor operator to the part of $u$ that depends only on $V(\la)$, i.e.
    \begin{align}
      u[\G\lineco \g] = u[\la\lineco\g] u[\G\lineco\la]. 
    \end{align}
    and in scaling
    \begin{align}
    \uvs_{x_\la} & \left( u[\g] \left(  - t^{d(\g)}_{x_\g\mid x_{\ol{\g}}} u[\G\lineco \g] \right) \right)                                                                                      \\
      & = \uvs_{x_\la}\left( u[\g] \right) + \uvs_{x_\la}\left( - t^{d(\g)}_{x_\g\mid x_{\ol{\g}}} u[\la\lineco \g] \right) + \uvs_{x_\la}\left( u[\G\lineco\la] \right)        \\
      & \leq \uvs_{x_\la}\left( u[\g] \right) + \uvs_{x_\la}\left( u[\la\lineco \g] \right) + \uvs_{x_\la}\left( u[\G\lineco\la] \right) \\
      & \leq \uvs_{x_\la}\left( u[\la] \right).                                                                         
    \end{align}
  \end{description}
  We turn to the second assertion. Since $u[\G]= u[\g] u[\G\lineco \g]$, we have
  \begin{align} 
    \uvs_{x_\g} & \left( \left( 1- t^{d(\g)}_{x_\g\mid x_{\ol{\g}}} \IP(\g) \right) u[\G] \right) \\
    & = \uvs_{x_\g} \left( u[\g] \left( 1- t^{d(\g)}_{x_\g\mid x_{\ol{\g}}} \right) u[\G\lineco\g] \right)                                  \\
    & \leq \uvs_{x_\g} \left( u[\g] \right) + \uvs_{x_\g} \left( u[\G\lineco\g] \right) - d(\g) - 1 \\
    & \leq \uvs_{x_\g} \left( u[\g] \right) - d(\g) - 1.
  \end{align}
  This concludes the proof.
\end{proof}
Note that we excluded the case of overlapping graphs $\la$ for the scaling and $\g$ for the Taylor surgery. The argument for this is given in the following. Let $\g,\la\subset\G$ and $\g\olap\la$ while maintaining that $\g$ and $\la$ are full vertex parts. Further consider $\g\in F$ and the scaling of $\la$. Then $\IP(\g)$ sorts out the line complement, so that we take $V(\G\lineco\g)$ which certainly has non-vanishing intersection with $V(\g)$. From this we single out the set $E'(\G\lineco\g)$, which is entirely in $\la$ and for each $e\in E'(\G\lineco\g)$ there exists a $v\in V(\g)$ such that $v\in\pa e$. \\
In the contraction of $\la$ to a point, edges of $E'(\G\lineco\g)$ are not collapsing to a point. 
The argument for this starts from the observation that the Taylor operator maps
\begin{align}
  x_v \mapsto x_{\ol \g} 
\end{align}
for $v\in V(\g)\cap V(\G\lineco\g)$ and, since $V(\g)\olap V(\la)$, we choose $v\in V(\g)\cap V(\la)$. Without loss of generality we contract $\la$ to the origin and set for each $w\in V(\la)$
\begin{align}
  x_w\mapsto \rh x_w 
\end{align}
with $\rh>0$. Then, in the contraction, we decompose $x_{\ol \g}$ accordingly
\begin{align}
  x_{\ol \g} & \bydef \sum_{v\in V(\g)} \xi_v x_v \\
        & =  \sum_{v\in V(\g)\setminus V(\la)} \xi_v x_v  + \sum_{w\in V(\g)\cap V(\la)} \xi_w x_w  
\end{align}
such that the scaled quantity is given by
\begin{align}
  x_{\ol \g,\rh} = \sum_{v\in V(\g)\setminus V(\la)} \xi_v x_v  + \rh \sum_{w\in V(\g)\cap V(\la)} \xi_v x_w 
\end{align}
and we obtain in the limit
\begin{align}
  \lim_{\rh\rightarrow 0} x_{\ol \g,\rh} = \sum_{v\in V(\g)\setminus V(\la)} \xi_v x_v \,.
\end{align}
Hence for each $e\in E'(\G\lineco\g)$, we have 
\begin{align}
  u[e](x_v,x_w) \mapsto u[e](x_{\ol \g},x_w) \rightarrow u[e](x_{\ol \g,\rh=0},0). 
\end{align}
Therefore, as long as the sum of vertices $v\in V(\g)\setminus V(\la)$ does not vanish, the elements of $E'(\G\lineco\g)$ do not collapse in the contraction and thus improve the scaling since we assumed positive scaling degree of every edge in $\G$.

In the next step, we prove that $Ru[\G]$ has the right scaling for any chosen $I$.
\begin{lemma}\label{le:recudim}
  Let $F$ be a saturated forest and $\g\in F$. Then
  \begin{align}
    \begin{array}{ll}
    \uvs_{I_\g}\left( R'_Fu[\g] \right) < d \lvert I_\g \rvert & \mbox{ for constant } \g \\
    \uvs_{I_\g}\left( R'_Fu[\g] \right) \leq \uvd_\g(u[\g]) + d\lvert I_\g \rvert & \mbox{ for integrated of variable } \g 
    \end{array}
  \end{align}
  provided that the same relations hold for any maximal element in $F$ wrt $\g$. 
\end{lemma}
\begin{proof}
  Let $\g\in F$ be constant. Then $\g$ has only variable or constant maximal subgraphs, where the variable subgraphs are in $H_F$. Hence we rewrite accordingly
  \begin{align}
    R'_Fu[\g] & = u[\g_1]...u[\g_a] \prod_{j=1}^a \ch(\g_j) u[\G\lineco \g_1...\g_a \mid_\g] \\
             & = u[\g_1]...u[\g_a] \prod_{i=1}^b (1-t^{d(\g_i)}_{x_{\g_i}\mid x_{\ol{\g}_i}}) 
             \prod_{j=b+1}^a (-t^{d(\g_j)}_{x_{\g_j}\mid x_{\ol{\g}_j}}) u[\G\lineco \g_1...\g_a \mid_\g]. 
  \end{align}
  Note that $u[\G\lineco \g_1...\g_a \mid_\g]$ contains only constant vertices except for those belonging to maximal subgraphs, which are set to the same point in the Taylor surgery for each subgraph, respectively. Then the scaling does not affect $u[\G\lineco \g_1...\g_a \mid_\g]$ since all maximal subgraphs are mutually disjoint and contain at least one constant vertex. Then we obtain only moments $(x-\ol x)^\al$ for $\g_1...\g_a$ in the scaling of $I_\g$. 
  We compute
  \begin{align}
    \uvs_{I_\g} & \left( R'_Fu[\g] \right)  
     = \uvs_{I_\g}( u[\g_1]...u[\g_a] \prod_{i=1}^b (1-t^{d(\g_i)}_{x_{\g_i}\mid x_{\ol{\g}_i}})
    \prod_{j=b+1}^a (-t^{d(\g_j)}_{x_{\g_j}\mid x_{\ol{\g}_j}}) u[\G\lineco \g_1...\g_a \mid_\g]) \\
    & \stackrel{\eqref{eq:LZUVSum}}{=} \sum_{k=1}^a \uvs_{I_{\g_k}}(u[\g_k]) + \uvs_{I_\g}( \prod_{i=1}^b (1-t^{d(\g_i)}_{x_{\g_i}\mid x_{\ol{\g}_i}})
    \prod_{j=b+1}^a (-t^{d(\g_j)}_{x_{\g_j}\mid x_{\ol{\g}_j}}) u[\G\lineco \g_1...\g_a \mid_\g]) ) \\
    \intertext{applying Lemma \ref{le:Taylor} and restricting to resulting moments}
    & \leq \sum_{k=1}^a \uvs_{I_{\g_k}}(u[\g_k]) + \uvs_{I_\g}\Big( \prod_{i=1}^b \sum_{\lvert \al_i \rvert=d(\g_i)+1} (x- x_{\ol \g_i})^{\al_i}\mid _{\g_i} \times \nonumber \\
    & \hspace{5.5cm} \prod_{j=b+1}^a \sum_{\lvert \be_j \rvert=0}^{d(\g_j)} -(x - x_{\ol \g_j})^{\be_j}\mid_{\g_j} \Big) \\
    \intertext{splitting the first sum}
    & = \sum_{i=1}^b \uvs_{I_{\g_i}}\Big(\sum_{\lvert \al_i \rvert=d(\g_i)+1} (x- x_{\ol \g_i})^{\al_i}\mid_{\g_i} u[\g_i]\Big) \nonumber \\
    & \hspace{3cm} + \sum_{j=b+1}^a \uvs_{I_{\g_j}}\Big(\sum_{\lvert \be_j \rvert=0}^{d(\g_j)} -(x - x_{\ol \g_j})^{\be_j}\mid_{\g_j} u[\g_j]\Big) \\
    \intertext{using Lemma \ref{le:BFscale} for the first term and Lemma \ref{le:uvstaylor} for the second term}
    & \leq \sum_{i=1}^b \uvs_{I_{\g_i}}(u[\g_i]) - d(\g_i) - 1  + \sum_{j=b+1}^a \uvs_{I_{\g_j}}(u[\g_j]) \\
    \intertext{applying the definition of UV-degree of divergence}
    & = \sum_{i=1}^b d \lvert I_{\g_i} \rvert + \underbrace{\uvd_{\g_i}(u[\g_i]) - d(\g_i) - 1}_{<0}  + \sum_{j=b+1}^a \underbrace{\uvs_{I_{\g_j}}(u[\g_j])}_{<d \lvert I_{\g_j} \rvert \mbox{ by hypothesis}} \\
    & < \sum_{i=1}^b d \lvert I_{\g_i} \rvert + \sum_{j=b+1}^a d \lvert I_{\g_j} \rvert = \sum_{i=1}^a d \lvert I_{\g_i} \rvert = d \lvert I_\g \rvert
  \end{align}
  Next let $\g\in F$ be variable or integrated. Then there are only variable or integrated maximal subgraphs by definition, but none of those are in $H_F$. We write
  \begin{align}
    R'_Fu[\g] & = u[\g_1]...u[\g_a] \prod_{j=1}^a \ch(\g_j) u[\G\lineco \g_1...\g_a \mid_\g]                                       \\
             & = u[\g_1]...u[\g_a] \prod_{j=1}^a (-t^{d(\g_j)}_{x_{\g_j}\mid x_{\ol{\g}_j}}) u[\G\lineco \g_1...\g_a \mid_\g]. 
  \end{align}
  Here, $u[\G\lineco \g_1...\g_a \mid_\g]$ carries integrated vertices and is involved in the scaling procedure. Therefore we keep the factor explicit in our computation. 
  We find
  \begin{align}
    \uvs_{I_\g} & \left( R'_Fu[\g] \right) = \uvs_{I_\g}\Big( u[\g_1]...u[\g_a] \prod_{j=1}^a (-t^{d(\g_j)}_{x_{\g_j}\mid x_{\ol{\g}_j}}) u[\G\lineco \g_1...\g_a \mid_\g] \Big) \\ 
     & \stackrel{\eqref{eq:LZUVSum}}{=} \sum_{i=1}^a \uvs_{I_{\g_i}}(u[\g_i])  \nonumber\\
     & \hspace{1cm} + \uvs_{I_\g}\Big( \prod_{j=1}^a \sum_{\lvert \al_j \rvert=0}^{d(\g_j)} \frac{(x-x_{\ol \g_j})^{\al_j}\mid_{\g_j}}{\al!} D^{\al_j}_{x_{\g_j}\mid x_{\ol \g_j}}  u[\G\lineco \g_1...\g_a \mid_\g] \Big) \\
     & \stackrel{\eqref{eq:LZUVProd}}{\leq} \sum_{i=1}^a \uvs_{I_{\g_i}}(u[\g_i]) \nonumber \\
     & \hspace{0.5cm} + \max_{\al_1...\al_a}\Big\{ \uvs_{I_\g}\Big( \prod_{j=1}^a (x-x_{\ol \g_j})^{\al_j}\mid_{\g_j} D^{\al_j}_{x_{\g_j}\mid x_{\ol \g_j}}  u[\G\lineco \g_1...\g_a \mid_\g] \Big) \Big\}  \,,          \\
     \intertext{where $D^{\al_j}_{x_{\g_j}\mid x_{\ol \g_j}}$ is differentiating wrt to $x_{\g_j}$ according to the multiindex $\al_j$, and after applying Lemma \ref{le:Taylor}}
     & \leq \sum_{i=1}^a \uvs_{I_{\g_i}}(u[\g_i]) + \uvs_{I_\g}( u[\G\lineco \g_1...\g_a \mid_\g] )  \\
     & = \uvs_{I_\g}\left( u[\g] \right) = \uvd_\g(u[\g]) + d \lvert I_\g \rvert          
  \end{align}
  This finishes the proof.
\end{proof}
Only the actual convergence of our method is left to show.
Recall that it is sufficient to show that the degree of divergence for each summand of the forest formula \eqref{eq:forestformula} is negative, after the reordering via saturation and for any integration set $I$. Due to the recursive structure of our approach, we have 
\begin{align}
  \uvs_I(R'_Su[\G])  \begin{cases}               
  < d \lvert I \rvert \mbox{ for } \G\notin S   \\
  \leq \uvd_\G(u[\G]) + d \lvert I \rvert \mbox{ for } \G\in S 
  \end{cases}                                       
\end{align}
where $R'_S$ refers to the $R$-operation with contributions coming from $\G$ excluded. For $\G\in S$, $\G\in H_S$ and thus $\chi(\G)=(1-t(\G))\IP(\G)$. Therefore we apply Lemma \ref{le:Taylor} and obtain
\begin{align}
  \uvs_I(Ru[\G]) < d \lvert I \rvert. 
\end{align}
This completes the proof of Theorem \ref{th:L1loc}.

\section{Proof of Theorem \ref{th:L1}}\label{se:L1}

We know from Theorem \ref{th:L1loc} that $Ru[\G]$ is locally integrable. Thus we focus on the infrared. The strategy to control the long range effects of Taylor operations is inspired by \cite{Lowenstein:1975ps} and based on establishing a relation of the long range behavior of the weight between the ``pure'' $u[\G]$ and the $R$-modified $Ru[\G]$. Clearly, $u[\G]$ is easier to control. Therefore we show that $Ru[\G]$ has better regularity for large arguments than $u[\G]$. In fact, we have to perform a slight modification of the latter since local integrability is not maintained in the transition from $Ru[\G]$ to $u[\G]$.

\subsection{Reduction}

The modification is modeled on the Calderon-Zygmund Lemma \cite[Theorem 4 of part I.3]{stein1970singular} and the associated Calderon-Zygmund decomposition, where the latter already assumes integrability, while we want to show integrability knowing only certain decay properties of the unmodified weight. 
But in fact, we know that $Ru[\G]$ is locally integrable. 
%Hence recall that we had to perform a reordering of the forest formula in regions $B_{\rh^*}(\bullet)$, i.e. mutually disjoint regions containing the graph diagonals of $Ru[\G]$ with respect to the chosen integration set $I$. 
Assume further that $Ru[\G]$ is integrable due to
\begin{align}\label{eq:IRAssumption}
  \irs_I(Ru[\G]) \geq \irs_I(u[\G])>0
\end{align}
for each $I\subset V(\G)$. 
Then we can introduce a constant $C>0$ and a set $O\subset \IR^{d \lvert I \rvert}$ such that $O=\bigcup_j Q_j$ consists of mutually disjoint open cubes $Q_j$ and $\lvert Ru[\G] \rvert \leq C$ almost everywhere on $\IR^{d \lvert I \rvert}\setminus O$. Additionally the following bounds hold for every cube $Q_j$.
\begin{align}
  C\leq \frac{1}{\m(Q_j)} \int_{Q_j} \lvert Ru[\G] \rvert d\m_I \leq 2^{d \lvert I \rvert} C
\end{align}
This enables us to perform the Calderon-Zygmund decomposition, i.e. we write $Ru[\G]=G[\G;I] + B[\G;I]$, where 
\begin{align}
  G[\G;I](x) = \begin{cases}
    Ru[\G](x) & x\in \IR^{d \lvert I \rvert}\setminus O \\
    \frac{1}{\m(Q_j)} \int_{Q_j} Ru[\G](x) d\m_I & x\in Q_j
  \end{cases}
\end{align}
is the ``good'' function and accordingly the ``bad'' function is given by \linebreak$B[\G;I](x) =0$ for $x\in \IR^{d \lvert I \rvert}\setminus O$ and $\int_{Q_j} B[\G;I]d\m_I = 0$ for each $Q_j$. Now note that
\begin{align}
  \|Ru[\G]\|_{L^1(\IR^{d \lvert I \rvert})} \leq \|G[\G;I]\|_{L^1(\IR^{d \lvert I \rvert})} + \| B[\G;I] \|_{L^1(\IR^{d \lvert I \rvert})}
\end{align}
and 
\begin{align}
  \|B[\G;I]\|_{L^1(\IR^{d \lvert I \rvert})} & = \int_{\IR^{d \lvert I \rvert}} \lvert B[\G;I](x)\rvert d\m_I = \sum_j \int_{Q_j} \lvert B[\G;I](x)\rvert d\m_I \\
  & \leq \sum_j \int_{Q_j} \big( \lvert Ru[\G](x)\rvert + \lvert G[\G;I](x)\rvert \big) d\m_I \\
  & \leq \sum_j 2 \int_{Q_j} \lvert Ru[\G](x)\rvert d\m_I < \infty,
\end{align}
where the last line follows from the definition of $G[\G;I]$ and the local integrability of $Ru[\G]$.
Therefore we have to show that the ``good'' function $G[\G;I]$ is absolutely integrable. We observe that $\lvert G[\G;I] \rvert \leq 2^{d \lvert I \rvert} C$ in every $Q_j$ and $\lvert G[\G;I] \rvert \leq C$ almost everywhere on $\IR^{d \lvert I \rvert}\setminus O$. In particular, we have $G[\G;I]=Ru[\G]$ for $\lvert x \rvert \rightarrow \infty$ so that it is sufficient to examine the scaling of $Ru[\G]$ for large arguments. 

Recall that the reordering was constructed for configurations where $Ru[\G]$ was singular. Hence we may work with
\begin{align}
  Ru[\G] \bydef \sum_{F\in\SF}\limits \prod_{\g\in F} (-t^{d(\g)}_{x_\g\mid x_{\ol{\g}}} \IP(\g)) u[\G].
\end{align}
But it is intuitively clear that the integrability will be determined by the ``worst'' summand in the forest formula, i.e. the summand with the slowest decay at infinity. Then we may restrict ourselves to any forest $F\in\SF$ and thus to
\begin{align}
  R_F u[\G] \bydef \prod_{\g\in F} (-t^{d(\g)}_{x_\g\mid x_{\ol{\g}}} \IP(\g)) u[\G].
\end{align}

\subsection{Recursion}

Analogously to the UV-case, we examine first the influence of moments and derivatives on the IR-scaling degree. 
\begin{lemma}\label{le:IRMomAndDev}
  Let $\al$ be a multiindex and $f\in C^\infty(\IR^n\setminus \OO)$. Then
  \begin{align}
    \irs_x(x^\al f) & \geq \irs_x(f)-\lvert\al\rvert \\
    \irs_x(\pa^\al f) & \geq \irs_x(f) + \lvert\al\rvert.
  \end{align}
\end{lemma}
\begin{proof}
  We compute
  \begin{align}
    \irs_x\left( x^\al \pa^\be u \right) & = \sup_{\ka\in\IR}\{\lim_{\La\rightarrow\infty} \La^\ka x^\al_\La \pa^\be_\La u_\La\} \\
    & = \sup_{\ka\in\IR}\{\lim_{\La\rightarrow\infty} \La^{\ka+\lvert\al\rvert-\lvert \be \rvert} x^\al \pa^\be u_\La\} \\
    & \geq \sup_{\ka\in\IR}\{\lim_{\La\rightarrow\infty} \La^{\ka} x^\al \pa^\be u_\La\} -\lvert\al\rvert + \lvert \be \rvert  \\
    & = \irs_x(u) -\lvert\al\rvert + \lvert \be \rvert.
  \end{align}
  For $\lvert \be \rvert=0$ we obtain the first assertion and for $\lvert\al\rvert=0$ the second statement follows.
\end{proof}  
Additionally we have \cite{Lowenstein:1975rg}
\begin{align}
  \irs\Big(\sum_j f_j\Big) & \geq \min_j \irs\left(f_j\right) \\
  \irs\Big(\prod_j f_j\Big) & = \sum_j \irs\left(f_j\right).
\end{align}
Coming back to the initial idea of the proof, we notice that due to the assumption in \eqref{eq:IRAssumption} we reduced the problem to the comparison of $R_F u[\G]$ and $u[\G]$ in terms of long range scaling properties. It is left to show that the assumption on the unmodified weight $u[\G]$ suffices to guarantee integrability of the $R$-modified weight $Ru[\G]$. We start by looking at the effect of Taylor operators on the IR-scaling degree, acting on sufficiently smooth functions.
\begin{lemma}\label{le:IRSonF}
  Let $f\in C^k(\IR^m\times \IR^n)$ and $k\geq d$. Then we have
  \begin{align}
    & \irs_x(t^d_{x\mid\ol x} f(x,y)) \geq \irs_x(f(x,y)) \\
    & \irs_y(t^d_{x\mid\ol x} f(x,y)) \geq \irs_y(f(x,y)) \\
    & \irs_{x,y}(t^d_{x\mid\ol x} f(x,y)) \geq \irs_{x,y}(f(x,y)) 
  \end{align}
\end{lemma}
\begin{proof}
  Regardless of the scaling variable, we may write
  \begin{align}
    \irs(t^d_{x\mid\ol x} f(x,y)) & = \irs \Big( \sum_{\lvert\al\rvert=0}^d \frac{(x-\ol x)^\al}{\al!} D^\al_{x\mid\ol x} f(x,y) \Big) \\  
    & \geq \min_\al \irs \left((x-\ol x)^\al D^\al_{x\mid\ol x} f(x,y) \right).
  \end{align}
  For a better distinction, we assume two different parameters $\al$ for the moments and $\be$ for the derivatives so that we find ourselves in the situation of Lemma \ref{le:IRMomAndDev}. Then we have
  \begin{align}
    \irs_{x/(x,y)} \left((x-\ol x)^\al D^\be_{x\mid\ol x} f(x,y) \right) & = \irs_{x/(x,y)} \left( f(x,y) \right) -\lvert\al\rvert +\lvert \be \rvert \\
    \intertext{and}
    \irs_y \left((x-\ol x)^\al D^\be_{x\mid\ol x} f(x,y) \right) & = \irs_y \left( f(x,y) \right) + \lvert \be \rvert.
  \end{align}
  Setting $\al=\be$ and minimizing over this parameter, the three assertions follow.
\end{proof}
Next, let us analyze the effect of the Taylor operations of a subgraph $\g\subseteq \G$ in various constellations regarding the IR-scaling of a subgraph $\la\subseteq \G$. Due to our simplification in the beginning of the proof, we only have to consider Taylor polynomials. 
\begin{lemma}\label{le:IR1Tay}
  We have
  \begin{align}
    \irs_{x_\la}\left( t^{d(\g)}_{x_\g\mid x_{\ol \g}} \IP(\g) u[\G] \right)\geq \irs_{x_\la}(u[\la])
  \end{align}
  for all $\la,\g\subseteq\G$.
\end{lemma}
\begin{proof}
  We start by working out the involved lines in both the scaling and the Taylor operation and recall that a subgraph $\dot\g$ is the vertex set $V(\g)$ together with all incident edges to elements in it. 
  We observe that they must be in $\dot\g\cap\dot\la$ for the scaling and a subset of $\dot\g\lineco\g$ for the Taylor operation. For any $e\in E(\dot\la\cap\dot\g\lineco\g)$ it holds
  \begin{align}
    t^{d(\g)}_{x_\g\mid x_{\ol \g}} u[\dot\g\lineco\g] & = \sum_{\lvert\al\rvert=0}^{d(\g)} \frac{(x-\ol x)^\al}{\al!} D^\al_{x\mid\ol x}u[\dot\g\lineco\g] \\
    & = \sum_{a+\lvert\al\rvert=0}^{d(\g)} \frac{(x_{s(e)}-\ol x)^a}{a!} (D^\al_{x_{s(e)}\mid\ol x}u[e]) 
     \frac{(x-\ol x)^\al}{\al!} D^\al_{x\mid\ol x}u[\dot\g\lineco\g\lineco e].
  \end{align}
  Due to our choice, we know further that either $s(e)\in V(\la)$, $t(e)\in V(\la)$, or both $s(e), t(e)\in V(\la)$. But this matches exactly the relations of Lemma \ref{le:IRSonF} and thus the assertion follows.
\end{proof}
The last step towards the comparison of $R$-modified and unmodified weight is the analysis of the recursive action of the $R$-operation for any forest $F\in\SF$.
\begin{lemma}\label{le:IRRec}
  Let $F\in\SF$, $\g\in F$ and $I\subset V(\G)$. Then 
  \begin{align}
    \irs_{I_\g}(R'_Fu[\G\mid_\g])\geq \irs_{I_\g}(u[\g])
  \end{align}
  provided the same relations hold for any maximal subgraph of $\g$ in $F$.
\end{lemma}
\begin{proof}
  Suppose that $\g_1,...,\g_a\in F$ are maximal subgraphs of $\g\subseteq\G$. Then we write
  \begin{align}
    R'_F u[\G\mid_\g] = u[\G\mid_{\g_1}] ... u[\G\mid_{\g_a}] \prod_{j=1}^a (-t^{d(\g_j)}_{x_{\g_j}\mid x_{\ol \g_j}}) u[\G\lineco(\g_1...\g_a)\mid_\g].
  \end{align}
  Note that this is possible since we consider only forests $F$ before the saturation. Choosing some $I\subset V(\G)$, the IR-scaling with respect to the intersection $I\cap V(\g)= I_\g$ gives 
  \begin{align}
    \irs_{I_\g}(R'_F u[\G\mid_\g]) & = \sum_{j=1}^a \irs_{I_\g}(u[\G \mid_{\g_j}]) 
    + \irs_{I_\g}\left( \prod_{j=1}^a (-t^{d(\g_j)}_{x_{\g_j}\mid x_{\ol \g_j}}) u[\G\lineco(\g_1...\g_a)\mid_\g] \right) \\
    & \geq \sum_{j=1}^a \irs_{I_\g}(u[\G\mid_{\g_j}]) + \irs_{I_\g}\left( u[\G\lineco(\g_1...\g_a)\mid_\g] \right),
  \end{align}
  where the inequality follows from the application of Lemma \ref{le:IR1Tay}. We observe immediately that under the hypothesis 
  \begin{align}\label{eq:IRRecursiveStep}
    \irs_{I_\g}(R'_F u[\G \mid_{\g_j}]) \geq \irs_{I_\g}(u[\g_j])
  \end{align}
  for all maximal subgraphs, it follows that
  \begin{align}\label{eq:IRRecursive}
    \irs_{I_\g}(R'_F u[\G\mid_\g]) \geq \irs_{I_\g}(u[\g]).
  \end{align}
  That this hypothesis is sensible may best be observed by starting with $\g$ chosen to be minimal, i.e. there exist no subgraphs in $\g$ which are also renormalization parts, i.e.\ subgraphs with non-negative UV-degree of divergence. 
  In that case,
  \begin{align}
    R'_F u[\G\mid_\g] = u[\g]
  \end{align}
  and thus \eqref{eq:IRRecursive} is fulfilled for all minimal graphs in $F$. The next step is obvious. Take any subgraph $\g$ such that all its maximal subgraphs are minimal in $F$. Then assumption \eqref{eq:IRRecursiveStep} holds by the previous step. 
\end{proof}
With this, we obtain for every normal forest $F$
\begin{align}
  \irs_I(R_F u[\G]) \geq \irs_I(u[\G])
\end{align}
and for every full forest $F$
\begin{align}
  \irs_I(t^{d(\G)}_{x_\g\mid x_{\ol \g}} \IP(\G) R'_F u[\G]) \geq \irs_I(t^{d(\G)}_{x_\g\mid x_{\ol \g}} \IP(\G) u[\G]] \geq \irs_I[u[\G]),
\end{align}
where we used Lemma \ref{le:IR1Tay} in the last inequality. Summing over all forests we arrive at 
\begin{align}
  \irs_I(Ru[\G]) \geq \min_{F\in\SF} \irs_I(R_F u[\G]) \geq \irs_I(u[\G]).
\end{align}
Since we assumed that $\ird_I(u[\g])>0$ for all $\g\subseteq\G$, we know for the $R$-modified weight that $\ird_I(Ru[\G])>0$. Additionally, we have $\uvd_I(Ru[\G])<0$ by Theorem \ref{th:L1loc} so that $Ru[\G]$ is integrable over I. 
This completes the proof of Theorem \ref{th:L1}.

\section{Conclusion}\label{se:conlusion}

In the present paper we propose a BPHZ prescription in configuration space.
In order to formulate the problem, we applied formally the Fourier transformation to Feynman integrals after analyzing their components.
With the introduction of suitable Taylor subtractions about a weighted center of mass, we show that the graph weight is locally integrable.
For a quantum field theory, where the graph weight is a distribution kernel, this implies that the domain of the distribution can be extended to the whole space.
Therefore the result may be viewed as a basis for a formulation of BPHZ renormalization for quantum field theories on curved spacetimes \cite{Pottel:2017bb}.
In order to transfer the current proposal to a renormalization scheme, the freedom, introduced by the Taylor subtractions, has to be characterized.
Then locality, unitarity and covariance can be checked by proving the equivalence to the approach in \cite{Brunetti:1999jn,Hollands:2001nf,Hollands:2001fb}.\\
Furthermore, we characterized decay properties of graph weights at large distances and showed that the weights are absolutely integrable for certain algebraic decays.
This has important implications for the treatment of quantum field theories with massless fields.
In comparison to the BPHZ momentum space prescription, where Taylor subtraction at vanishing external momentum introduce new, unphysical IR-divergencies, which are cured by introducing an auxiliary mass term and further subtractions \cite{Lowenstein:1975ps}, the $R$-operation remains unchanged in our version.
Therefore massive and massless quantum fields can be treated on equal footing.
Additionally, our result ensures the existence of a constant coupling limit in a naive sense if a quantum field theory is initially formulated with local coupling functions.
In this regard, an interesting extension of the current work would be the study of bounds on the renormalized Feynman graphs \cite{Feldman:1985th}.
Furthermore, our results should allow for a formulation of the short distance expansion for various quantum fields \cite{Zimmermann:1972tv} directly in configuration space.

\subsection*{Acknowledgments}

The author would like to thank Klaus Sibold for numerous discussions. 
Parts of this work have been concluded during a stay at Leipzig University and the Max Planck Institute for Mathematics in the Sciences.
Their hospitality and the financial support by the International Max Planck Research School (IMPRS) ``Mathematics in the Sciences'' is gratefully acknowledged. 

%% Bib
\bibliographystyle{./halpha} 
\bibliography{./bphzl}

\end{document}